\documentclass[a4paper,twocolumn]{esapub2005} 
\usepackage{times,graphicx}
\usepackage{natbib} 
 
\title{Recent Radio Monitoring of Microquasars with RATAN-600 Radio Telescope}
\author{S. A. Trushkin}
\affil{Special astrophysical observatory RAS, Nizhnij Arkhyz, 369167, Russia, satr@sao.ru}
\author{N. N. Bursov}
\affil{Special astrophysical observatory RAS, Nizhnij Arkhyz, 369167, Russia, nnb@sao.ru}
\author{T. Kotani}
\affil{Tokyo Tech, 2-12-1 O-okayama, Tokyo 152-8551, Japan, kotani@hp.phys.titech.ac.jp}
\author{N. A. Nizhelskij}
\affil{Special astrophysical observatory RAS, Nizhnij Arkhyz, 369167, Russia, nizh@sao.ru}
\author{M. Namiki}
\affil{Osaka University, 1-1 Machikaneyama, Toyonaka, Osaka 560-0043, Japan, namiki@ess.sci.osaka-u.ac.jp}
\author{M. Tsuboi}
\affil{Nobeyama Radio observatory, Minamimaki, Minamisaku, Nagano, 384-1305, Japan, tsuboi@nro.nao.ac.jp}
\author{P. A. Voitsik}
\affil{Moscow State University by M.V. Lomonosov, Moscow GSP-2, 119992, Russia, voitsik@sai.msu.ru}
 
\begin{document} 
\keywords{microquasars; radio emission; X-rays; monitoring}
\maketitle 
 
\begin{abstract} 
We report about the multi-frequency (1-30 GHz) daily monitoring of the
radio flux variability of the three microquasars: SS433, GRS1915+105 and
Cyg X-3 during the  period from September 2005 to May 2006.

1. We detected clear correlation of the flaring radio fluxes and
X-rays 'spikes' at 2-12 keV emission detected in
RXTE ASM from GRS1915+105 during eight relatively bright (200-600 mJy)
radio flares in October 2005. The 1-22 GHz spectra of these
flares in maximum were optically thick at frequencies lower 2.3 GHz and
optically thin at the higher frequencies.
During the radio flares the spectra of the X-ray spikes become
softer than those of the quiescent phase.
Thus these data indicated the transitions from very high/hard states
to high/soft ones during which massive ejections are probably happened.
These ejections are visible as the detected radio flares.

2. After of the quiescent radio emission we have detected a
drop down of the fluxes ($\sim$20 mJy) from Cyg X-3.
That is a sign of the following bright flare.
Indeed such a 1~Jy-flare was detected on 3 February 2006 after 18 days of
the quenched radio emission.
The daily spectra of the flare in the maximum was flat
from 1 to 100 GHz, using the quasi-simultaneous observations at 109
GHz with RT45m telescope and millimeter array (NMA) of Nobeyama Radio
Observatory in Japan. The several bright
radio flaring events (1-10 Jy) followed during this state of very
variable and intensive 1-12 keV X-ray emission ($\sim$0.5 Crab),
which being monitored in RXTE ASM program.
We discussed the various spectral and temporal characteristics of
the detected 180-day light curves from three microquasars in comparison
with Rossi XTE ASM data.
\end{abstract} 
 
\section{Introduction} 

Collimated high-velocity outflows of accreting matter in the narrow cones
of the two-side relativistic jets, ejected from polar regions of accretion
disks around black holes or neutron stars in the microquasars, are the
effective sources of variable synchrotron emission in distinct clouds
contained magnetic fields and energetic electrons.
Only 15 microquasars are now detected in the Milky Way in the sample
of 350 X-ray binaries.
The ballistic tracks of these clouds (blobs) are directly visible as radio
jets in VLA and VLBI maps of SS433, GRS 1915+105 and Cyg X-3.
The time and frequency dependences in the light curves are the key
for clear understanding and good probe test for models of the physical
processes in cosmic jets. A comparison of the radio and X-ray data
allow us to provide detailed studies.
We have carried out the 250-day monitoring of the microquasars
Cyg X-3, GRS~1915+105, and SS433, with the RATAN-600 radio telescope
at 1-30 GHz from September 2005 to May 2006.

\begin{figure}
\centering
\includegraphics[width=0.9\linewidth]{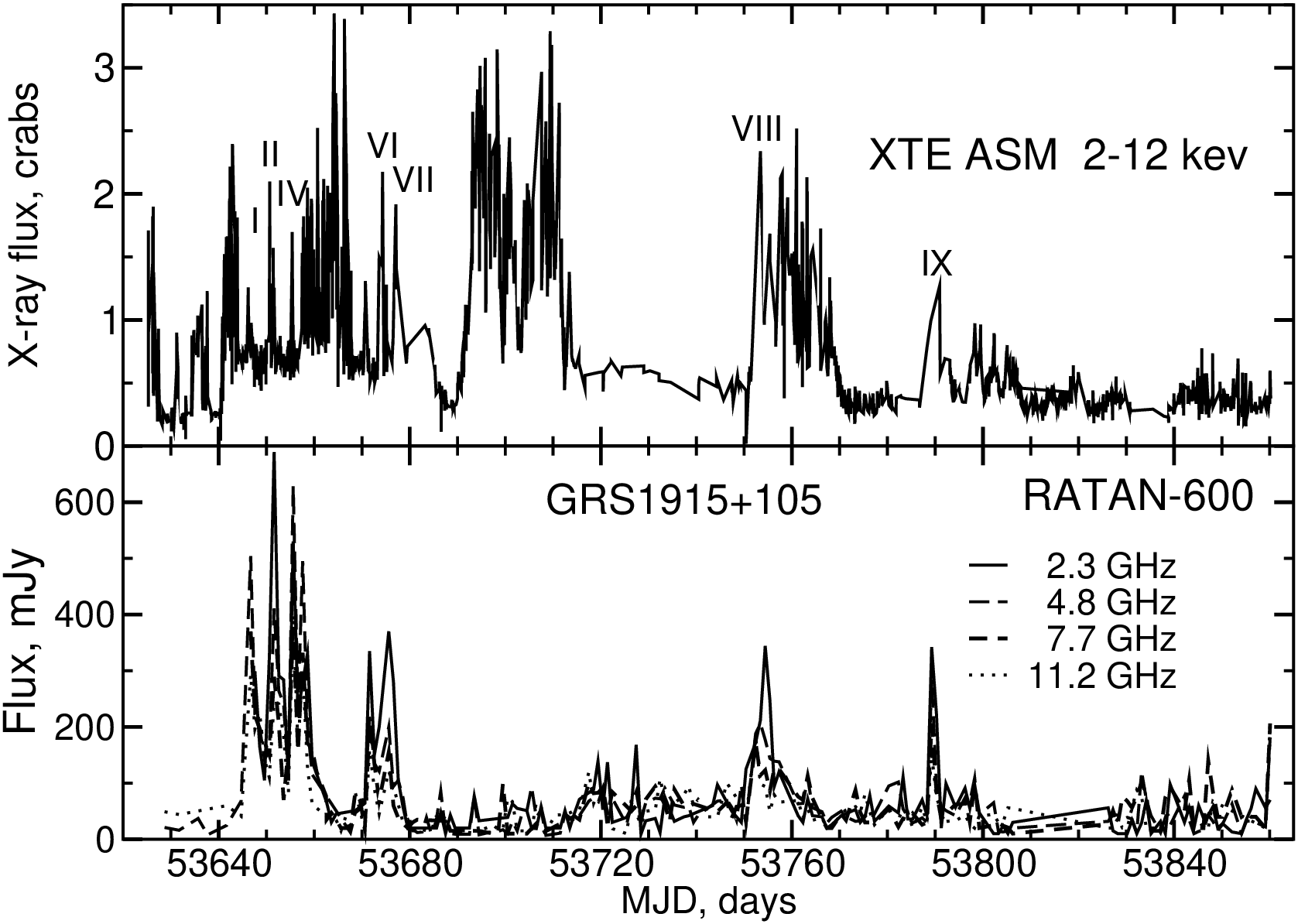}
\caption{Light curves of GRS1915+105 at radio frequencies and at 2-12 keV
from September 2005 to March 2006.}
\label{fig:1}
\end{figure}

\section{Observations}

We have carried out the 250-day almost daily monitoring observations
of the microquasars Cyg X-3, GRS~1915+10, SS433, with RATAN-600 radio
telescope at 1-22 GHz from September 2005 to May 2006.
The measured multi-frequency light curves
can be directly compared with series of the X-ray observatory
RXTE \citet{lev96}.
We have used a standard continuum radiometer complex. The
receivers at 3.9, 7.7, 11.2, and 21.7 GHz were equipped
with closed-cycle cryogenic systems, which lowered the temperature of the
first amplifiers (HEMT) to 15-20 K. Low-noise transistor amplifiers were
installed in the 0.98- and 2.3-GHz radiometers. Table 1
presents the current mean sensitivity of the RATAN-600 telescope
for a single transit of a source through the fixed antenna beam.
The observations were made using the `Northern sector' antenna of
RATAN-600 radio telescope at the upper culmination of the sources.

The flux densities of the sources at all six frequencies
were measured in a single observation. Note that the
resolution of the telescope was quite sufficient to reliably distinguish
SS433, GRS~1915+105 and Cyg X-3 against the Galactic background.
Although interference sometimes prevented realization of the maximum
sensitivity of the radiometers, daily observations of reference sources
indicate that the error in the flux density measurements did
not exceed 5-10\% at 2.3, 3.9, and 11.2 GHz and 10-15\% at 21.7 GHz.

\begin{table}
\begin{center}
\caption{Sensitivity of the RATAN-600 telescope.}\vspace{1em}
\renewcommand{\arraystretch}{1.2}
\begin{tabular}{llllll}
\hline
 $\lambda$, cm  31  & 13.0 &  ~7.6  &  ~3.9 &  ~2.7 & ~1.38 \\
 $\nu$, GHz     1.0 & ~2.3 &  ~3.9  &  ~7.7 &  11.2 & 21.7  \\
 $\Delta$S,mJy  30  & 10   &   3    &  10   &  10   & 15    \\
\hline
\end{tabular}
\label{tab:1}
\end{center}
\end{table}

The flux density calibration was performed using observations of 3C286
(1328+30), PKS~1345+12 and NGC7027 (2105+42). We have controlled the antenna
gain with thermal source (HII region) 1850--00 in daily observation also.
We took the fluxes for these sources from \citet{ali85},
which, in turn, were consistent with the primary radio astronomy flux scale
of \citet{baars77} and with the new flux measurements of \citet{ott94}.
The reference source fluxes adopted for this observation cycle are
presented in Table 2.

{\small
\begin{table}
\begin{center}
\caption{Flux densities for calibration sources, Jy}
\begin{tabular}{llllllll}
\hline
 Source   & \multicolumn{6}{c}{Frequency, GHz }          \\
\cline{1-6}
 name     &   1.0  &   2.3   &    4.8  &    7.7   &  11.2  &  21.7   \\
\hline
 1331+30  &  17.49 &  11.5   &    7.22  &   5.52  &   4.25 &   2.5   \\
 1345+12  &   6.5  &   4.26  &    2.95  &   2.25  &   1.82 &   0.94  \\
 1850-00  &    -   &   2.27  &    3.33  &   3.88  &   4.19 &   4.49  \\
 2105+42  &   0.95 &   3.04  &    5.05  &   5.86  &   6.10 &   5.71  \\
\hline
\end{tabular}
\label{tab:2}
\end{center}
\end{table}
}

The recording of the data and the preliminary reduction and storage
on a personal computer were carried out in the data collection package.
The data reduction of the data FITS-files included background removal,
convolution with the antenna beam, and Gaussian analysis.
In this way, instantaneous spectra of the microquasars were constructed
for each day from the measurements at the four-six frequencies.

\begin{figure}
\centering
\includegraphics[width=0.9\linewidth]{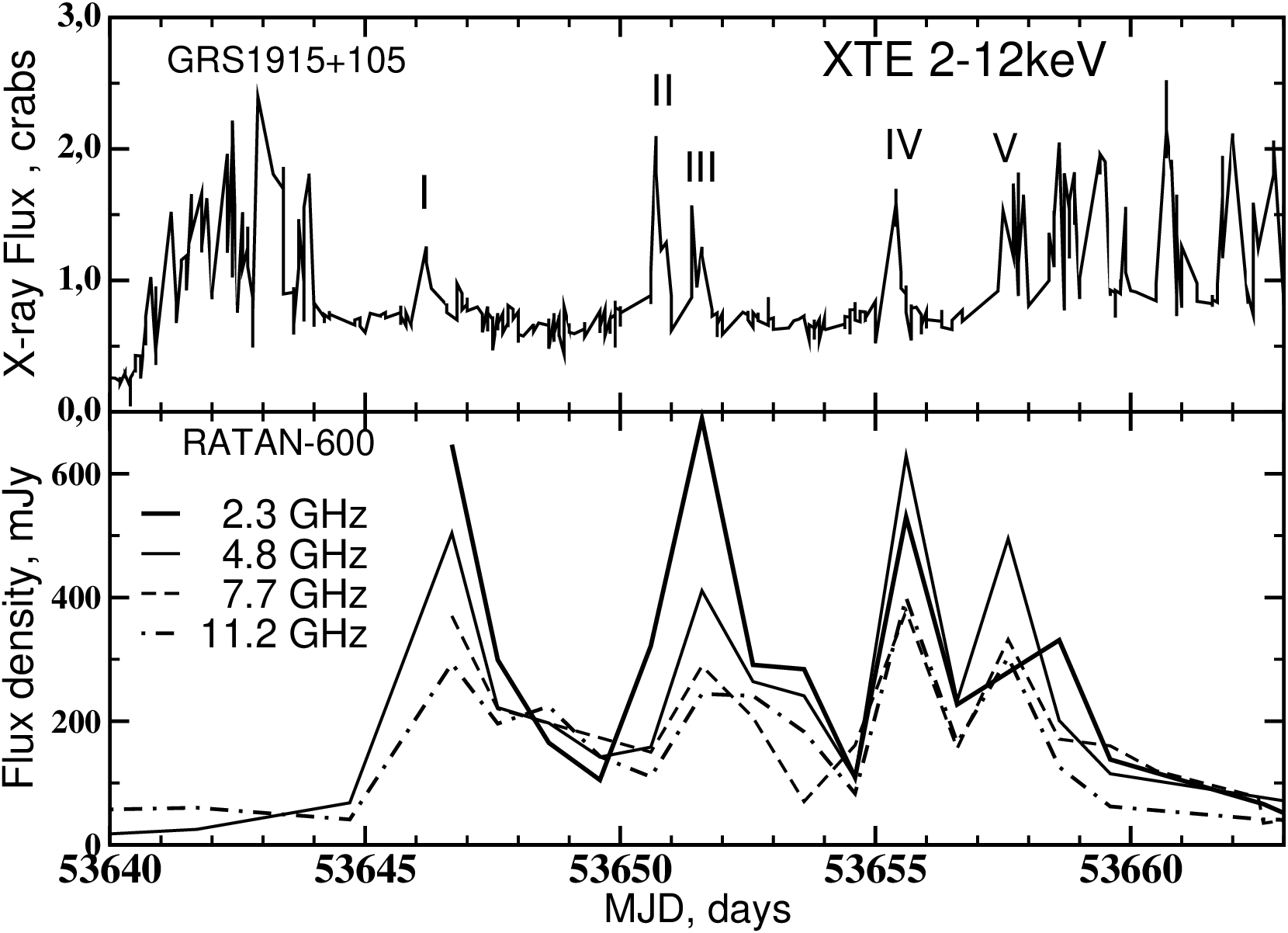}
\caption{The radio and X-ray  light curves of GRS1915+105 in October 2005.}
\label{fig:2}
\end{figure}

\section{Results}
\subsection{GRS 1915+105: X-ray/radio correlation}

The X-ray transient source GRS 1910+105 was discovered  by
\citet{CT92} with WATCH instrument on board GRANAT. In 1994,
a superluminal motion of the jet had been detected from GRS 1915+105
(\citet{mirabel94}), since then a new class of astrophysical
objects 'Microquasars' was established.

\begin{figure}
\centering
\includegraphics[width=0.9\linewidth]{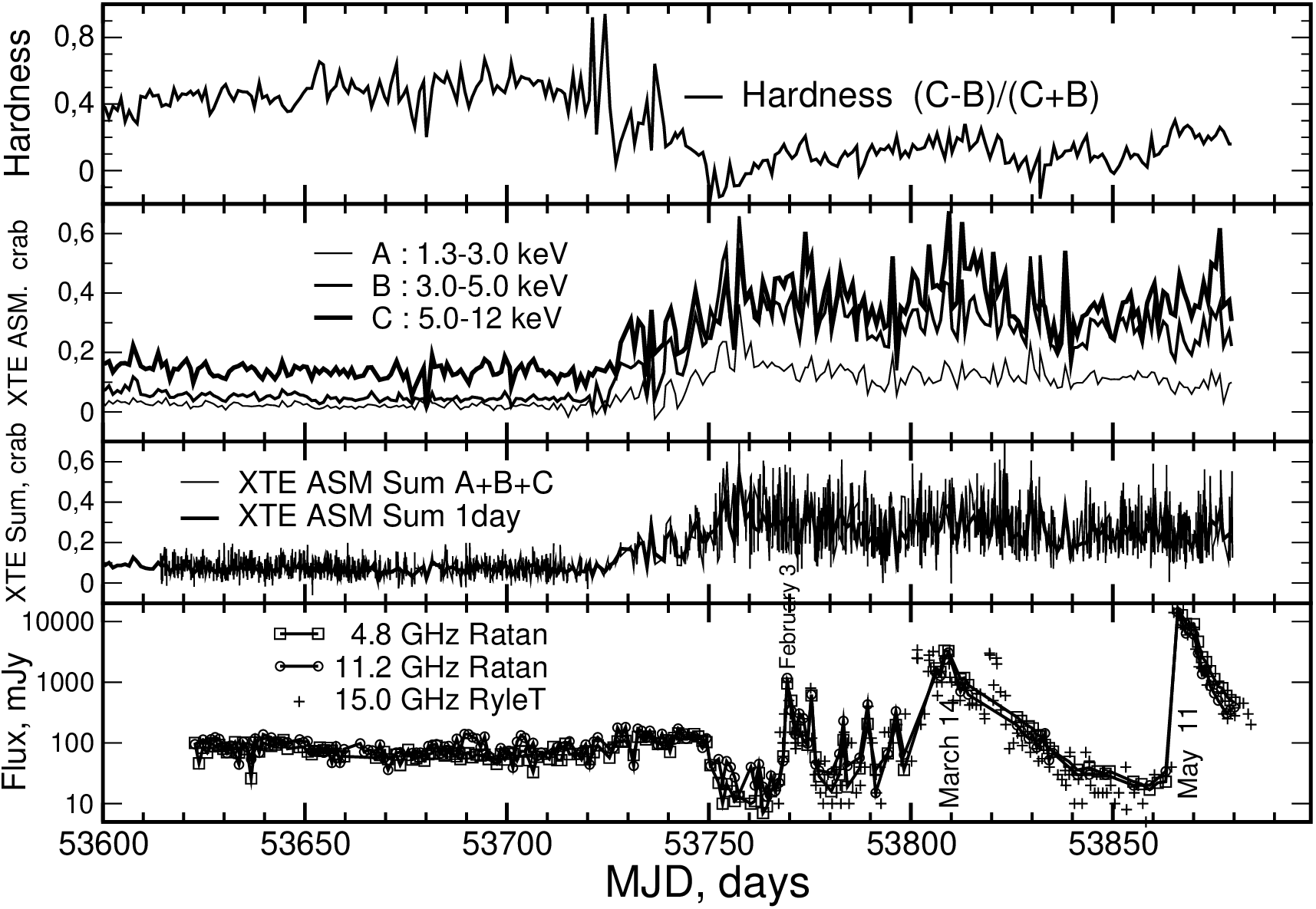}
\caption{
The RATAN and RXTE ASM light curves of Cyg X-3 from September 2005 to
May 2006.}
\end{figure}

Many X-ray observations of GRS 1910+105 revealed remarkable QPOs which
are believed to arise in the accretion disk around a black hole.  On the
other hand, we are far form the full understanding of the jet phenomena.
To interpret the X-ray data correctly, long-term radio monitoring data
are desirable. By reference to radio data, the activity and state of the
source can be diagnosed. Furthermore, the massive jet ejection events,
by which the source was recognized as a microquasar in the first place,
can be predicted by means of radio monitoring. A massive jet ejection
event from another microquasar, SS 433, was observed with RXTE with the
help of radio monitoring (\citet{kotani06}).

During the decay of the first flare (July 2000) \citet{fender02}
for the first time detected
the quasi-periodical oscillations with P = 30.87 minutes at two frequencies:
4800 and 8640 MHz. The linear polarization of the oscillations was measured
at a level 1-2 per cent with a flat spectrum.

In Fig.1 the radio and X-ray light curves are showed during the total set.
The nine radio flares have the counterparts in X-rays.
The radio spectrum was optically thin in the first two flares,
and optically thick in third one (Fig.2).
The profiles of the X-ray spikes during the radio flares
are clearly distinguishable from other spikes because of its shape, it
shows the fast-rise and the exponential-decay. The other X-ray spikes,
which reflect the activity of the accretion disk, show an irregular
pattern. During the radio flare, the spectra of the X-ray spikes become
softer than those of the quiescent phase, by a fraction of $\sim$30\%.

\begin{figure}
\centering
\includegraphics[width=0.9\linewidth]{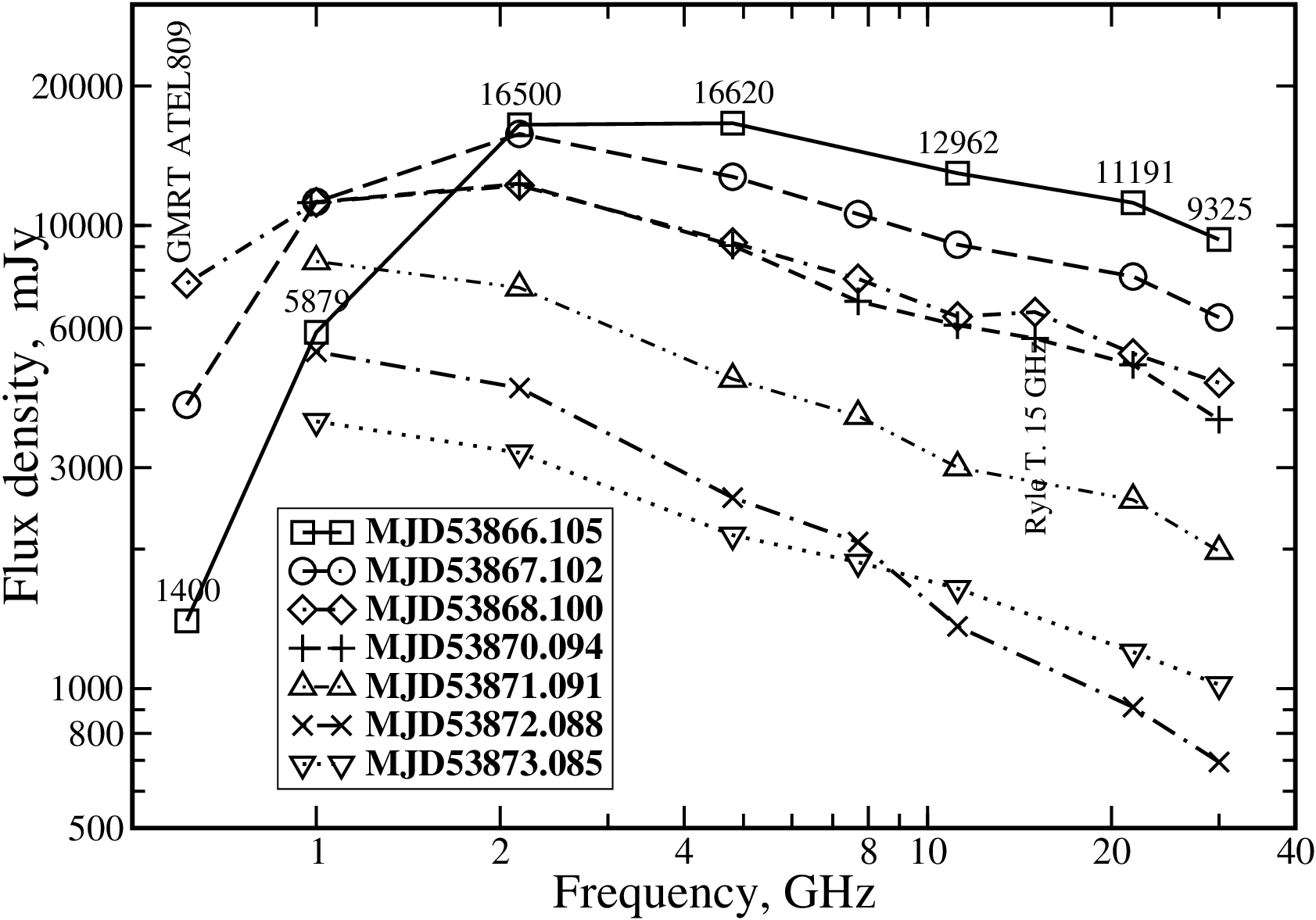}
\caption{The daily spectra of Cyg X-3 during flare in May 2006.}
\label{fig:3}
\end{figure}

\citet{miller06} have detected large-scale radio jet with VLBA mapping
during an X-ray and radio outburst on 23 February 2006 (MJD53789.258).
Then the optically thin flare with fluxes
340, 340, 342, 285, 206, and 153 mJy was detected at frequencies
1, 2.3, 4.8, 7.7, 11,2 and 21.7 GHz.

\subsection{Cyg X-3: new long active period }          

During ~100 days \mbox{Cyg~X-3} was in a quiescent state of $\sim$100 mJy (Fig.3).
In December 2005 its X-ray flux began to increase and radio flux at 2-11
GHz increased also. Then the flux density at 4.8 GHz of the source was
found to drop from 103 mJy on Jan 14.4 (UT) to 43 mJy on Jan 15.4 (UT),
and to 22 mJy on Jan 17.4 (UT). The source is known to exhibit the radio
flares typically with a few peaks exceeding 1-5 Jy following such quenched
state as Waltman \citet{waltman94} have showed in
the intensive monitoring of Cyg X-3 with the Green Bank
Interferometer at 2.25 and 8.7 GHz.
The source has been monitored from Jan 25 (UT) with the Nobeyama Radio
Observatory 45m Telescope (NRO45m Telescope),
the Nobeyama Millimeter Array (NMA), Yamaguchi-University 32-m
Radio Telescope (YRT32m), and Japanese VLBI Network telescopes.
On Feb 2.2 (UT), about 18 days after it entered the quenched state, the
rise of a first peak is detected with the NRO45m Telescope and YRT32m.
On Feb 3.2 (UT), the flux densities reached to the first peak at all the
sampling frequencies from 2.25 GHz to 110.10 GHz (\citet{tsuboi06}).
The spectrum in maximum (3 February) of the flare was flat
as measured by RATAN, NRO RT45m and NMA from 2 to 110 GHz.
The next peak of the active events on 10 February
reached the level of near 1 Jy again with a similar flat spectrum.
Then three short-time flare have happened during a week.
The flare on 18 February had the inverted spectrum with the same
spectral index $\alpha$=+0.75 from 2.3 to 22 GHz.

\begin{figure}
\centering
\includegraphics[width=0.9\linewidth]{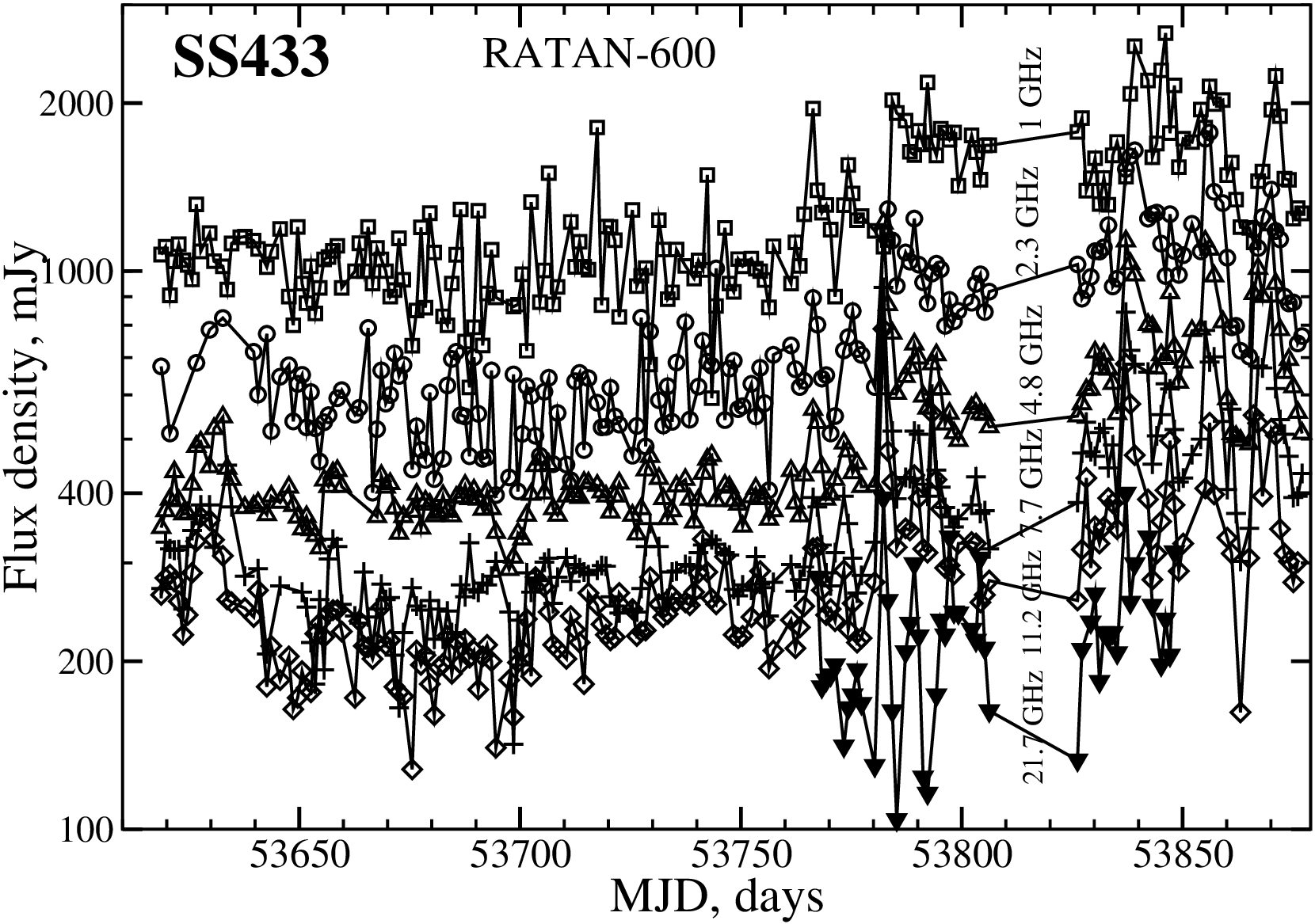}
\caption{The light curves of SS433 from September 2005 to May 2006.}
\label{fig:5}
\end{figure}

In the active period there were two powerful flares, March 14 to 3-5 Jy
and May 11 to 12-16 Jy at 2-30 GHz. In the May flare
fluxes have grown up by a factor $\sim$1000 during a one day.
Such powerful ejection of relativistic electrons
were detected with RATAN and by G. Pooley with Ryle telescope.

The change of the spectrum  during the flare on May 11-19
followed to model of single ejection of the relativistic electrons, moving
in thermal matter in the intensive WR-star wind.
It stays in optically thin mode at the higher frequencies,
meanwhile at lower frequency 614 MHz (\citet{pal06}, Fig.4).
Cyg X-3 was in hard absorption due to thermal electrons in stellar wind.

The decay of the synchrotron radio flares and spectral variability follow
prediction of a finite jets segment model, that was involved by
\citet{marti92}
and by  \citet{Hj00}  for modeling the flaring of Cyg X-3
and SAX J1819-25 respectively.
The typical `bubble' event follows by the optically
thin power-law decay $\mathbf{\propto\nu^{\alpha}\,t^{-\beta}}$, where
index $\mathbf\alpha$ varies from  --1 to 0, and
index $\mathbf\beta$ varies from 1 to 6 according to the dependence
from distance $r$: internal magnetic field H(r),
thermal N$_{\mathrm{th}}$(r), relativistic N$_{\mathrm{rel}}$(r)
electrons densities, and jet and expansion velocities
v$_{\mathrm{jet}}$(r), v$_{\mathrm{exp}}$(r).

\subsection{SS433: the light curves and spectra}

The first microquasar SS433, a bright variable emission star was
identified  with a rather bright compact radio
source 1909+048 located in the center of a supernova remnant W50.
In 1979 moving optical emission lines, Doppler-shifted due to precessing
mass outflows with 78000 km/s, were discovered in the spectrum of
SS433. At the same time in 1979 have discovered a unresolved
compact core and 1 arcsec long aligned jets in the MERLIN radio
image of SS433. At present such a structure in microquasars is commonly
named a radio jet. Different data do indicate a presence of a very
narrow (about 1$^o$) collimated beam at least in X-ray and optical
ranges. At present there is no doubt that SS433 is related to W50.
A distance of near 5 kpc was later determined
by different ways including the direct measurement of proper motions of
the jet radio components.

\citet{kotani06} detected the fast variation in the X-ray emission
of SS433 during the radio flares, probably even QPOs of 0.11 Hz.
Massive ejections during this active period could  be the reason of
such behavior.
In Fig.5 the daily RATAN light curves are shown.
Clearly the activity of SS433 began during the second half of the monitoring
set. Some flares happened just before and after the multi-band program
of the studies of SS433 in April 2006 (\citet{kotani07}).

In Fig.6 the light curve during the bright flare in February 2006 are
showed after subtracting a quiet spectrum $S_\nu[Jy] =1.1\,\nu^{-0.6}[GHz]$.
The delay of the maximum flux at 1 GHz is about 2 days and 1 day at 1 and 2.3
respectively relatively the maxima at the higher frequencies.
Below in Fig.6 the spectra of the flare during first three days.
We see the characteristic shift of turn-over of spectra to low frequency
during the flare, clearly indicating the decrease of the absorption with time.

\begin{figure}
\centering
\includegraphics[width=0.85\linewidth]{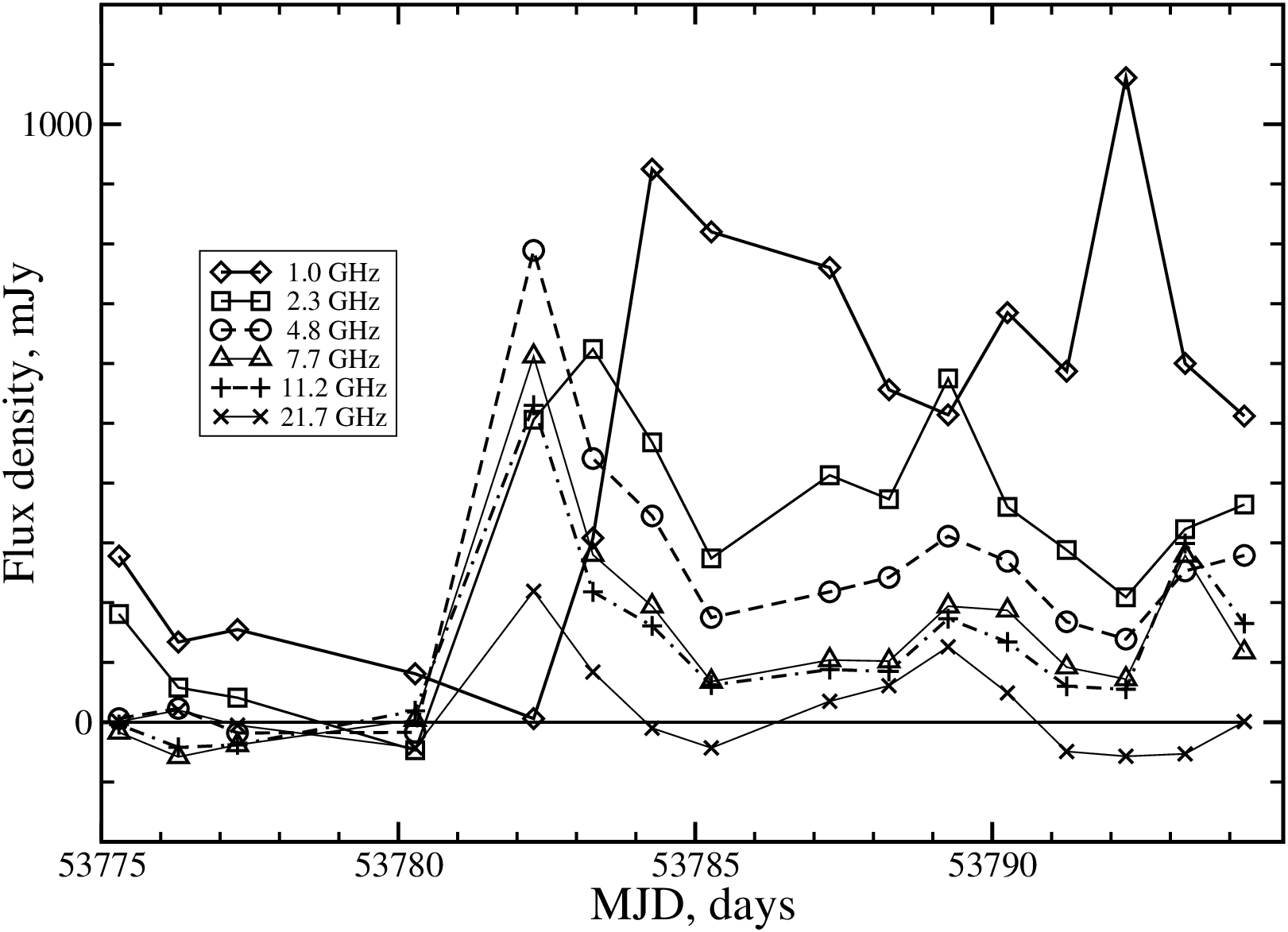}
\includegraphics[width=0.85\linewidth]{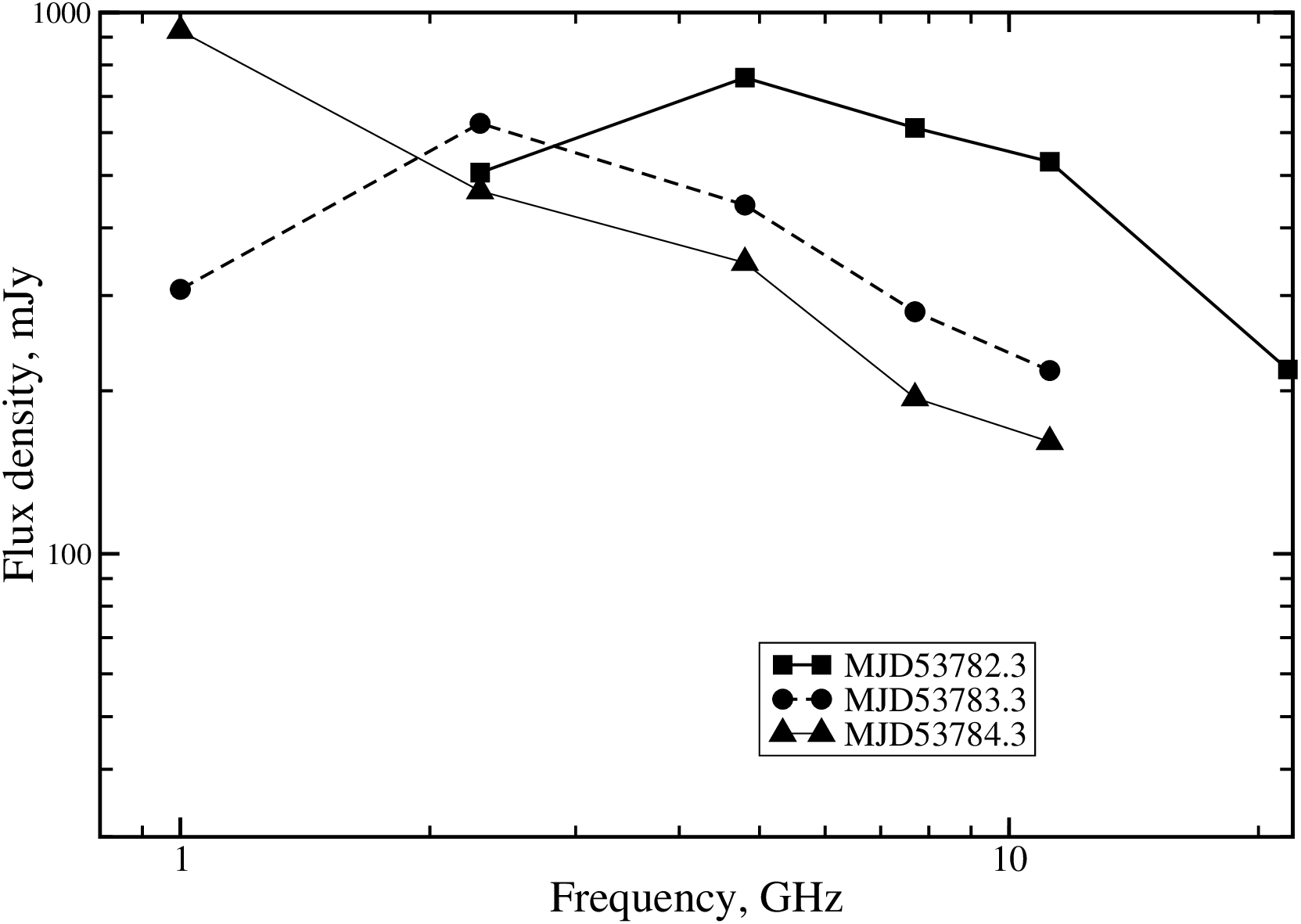}
\caption{The light curves after subtracting a quiet spectrum of SS433
in February 2006 and the flaring spectra  on 16-18 February 2006.}
\label{fig:6}
\end{figure}

\section*{Acknowledgments}
These studies were supported by Russian  Foundation  Base Research (RFBR)
grant N~05-02-17556 and  mutual RFBR and
Japan Society for the Promotion of Science (JSPS) grant N~05-02-19710.


\begin{thebibliography}{xx}
\bibitem[Levine et~al.(1996)]{lev96}
Levine A., Bradt H., Cui W., et al.,  1996, ApJ 469, L33

\bibitem[Aliakberov et~al.(1985)]{ali85}
Aliakberov K.D., Mingaliev M.G., Naugolnaya M.N., Trushkin S.A.,
et al., 1985, Izv. SAO 19, 60

\bibitem[Baars et al.(1977]{baars77}
Baars, J.W.M., Genzel, R., Pauliny-Toth, I.I.K., and Witzel, A.,
1977, A\&A  61, 99

\bibitem[Ott et~al.(1994)]{ott94}
Ott M., Witzel A., Quirrenbach A., et al., 1994, A\&A,  284, 331

\bibitem[Castro-Tirado et~al.(1992)]{CT92}
Castro-Tirado A. J., Brandt S., Lund N. 1992, IAUC \#5590

\bibitem[Mirabel F., Rodriguez, (1994)]{mirabel94}
Mirabel I. F., \& Rodriguez L. F. 1994, Nature 371, 46

\bibitem[Kotani et~al.(2006)]{kotani06}
Kotani T., Trushkin S. A., Valiullin R. K. et al., 2006,
ApJ 637, 486

\bibitem[Fender et~al.(2002)]{fender02}
Fender R.P., Rayner D.,  Trushkin S.A., et al. 2002
MNRAS 330, 212

\bibitem[Miller-Jones et~al.(2006)]{miller06}
Miller-Jones J.C. A., Rupen M.P., Trushkin, et al. 2006,
ATel \# 758, 1

\bibitem[Waltman(1994)]{waltman94}
Waltman E. B., Fiedler R.L., Johnston K. L., Ghigo F. D.
1994, AJ, 108, 179

\bibitem[Tsuboi et~al.(2006)]{tsuboi06}
Tsuboi M., Kuno N., Umemoto T., Sawada T., et al., 2006,
Kotani T., Kawai N.
ATel \#727, 1
 
\bibitem[Pal et~al.(2006)]{pal06}
Pal S., Ishwara-Chandra C. H.,  Pramesh A.
2006, ATel \#809

\bibitem[Marti et~al.(1992)]{marti92}
Marti J.,  Paredes J.M.,  Estalella R., 1992,
A\&A 258, 309

\bibitem[Hjellming et~al.(2000)]{Hj00}
Hjellming R. M., Rupen M. P., Hunstead R. W., et al.,
2000,  ApJ 544, 977

\bibitem[Kotani et~al.(2007)]{kotani07}
Kotani S., et al., 2007, VI microquasars workshop, Komo, Italy,
(in preparation)


\end{thebibliography}
\end{document}